# Integrating Identity-based Cryptography in IMS Service Authentication

Mohamed Abid[1], Songbo Song[2], Hassnaa Moustafa[2] and Hossam Afifi[1]

[1]Telecom & Management SudParis ,Evry, France
{Mohamed.Abid, Hossam.Afifi}@it-sudparis.eu
[2]Telecom R&D (Orange Labs), Issy Les Moulineaux, France
{Songbo.Song, Hassnaa.Moustafa}@orange-ftgroup.com

## Abstract

*Nowadays, the IP Multimedia Subsystem (IMS) is a promising research field. Many ongoing works related to the security and the performances of its employment are presented to the research community. Although, the security and data privacy aspects are very important in the IMS global objectives, they observe little attention so far. Secure access to multimedia services is based on SIP and HTTP digest on top of IMS architecture. The standard deploys AKA-MD5 for the terminal authentication. The third Generation Partnership Project (3GPP) provided Generic Bootstrapping Architecture (GBA) to authenticate the subscriber before accessing multimedia services over HTTP. In this paper, we propose a new IMS Service Authentication scheme using Identity Based cryptography (IBC). This new scheme will lead to better performances when there are simultaneous authentication requests using Identity-based Batch Verification. We analyzed the security of our new protocol and we presented a performance evaluation of its cryptographic operations.*

## 1. Introduction

The IP Multimedia Subsystem (IMS) [1], is an architectural framework for delivering Internet Protocol (IP) multimedia services. It was originally designed by the wireless standards body 3rd Generation Partnership Project (3GPP), as a part of the vision for evolving mobile networks beyond Global System for Mobile Communications (GSM). The IMS provides multimedia services (such as Voice over IP (VoIP), video conferencing, presence, push-to-talk etc.) on top of all IP networks as well as NGN (Next Generation Networks).

IMS provides a unique architecture for authentication and accounting of different services. Each subscriber uses an ISIM (IMS-SIM) card with a stored secret key to be able to authenticate to the IMS network and to be able to access the IMS services. One requirement provided in IMS, is the tight attachment of the subscriber authentication to the User Equipment UE since the ISIM card participates in the Authentication and key Agreement (AKA) [2]. Therefore, the keys are not generated from the user Identity but are randomly chosen by the Home Subscriber Server (HSS).

Consequently, IMS authentication falls short in one hand to realize authentication in a personalized manner, which is an important prerequisite in new services such as social internet ones. Moreover, using AKA in IMS proved to have some weakness, like short key for cryptographic purposes [3], [4]. Many solutions are provided to strengthen IMS security. Wu et al [5] define a new AKA based on Elliptic Curve Cryptography (ECC) and Huang et al [6] define a new AKA called one pass AKA for UMTS. Furthermore, Ring et al [7] tried to design a new AKA mechanism for [Session Initiation Protocol](#) (SIP) using Identity Based Cryptography (IBC) [8]. However, these works were only done for the subscriber authentication with nothing special on service authentication.

In this paper, we propose a new authentication scheme for services authentication in IMS. Our proposed scheme is an Identity based authentication type which in turn allows users' identification with a certain sort of personalization. The key advantage of IBC lies in its simplicity and robustness. Moreover, our solution proposes an Identity based batch verification which will decrease the verification time for simultaneous clients requesting service authentication.

The remainder of this paper is organized as follows: Section 2 describes the IMS service authentication process. Section 3 gives an overview on the IBC mechanism. In Section 4, we present our novel solution, and in Sections 5, we analyse its security. In Section 6, we briefly discuss the performance analysis that we carried out and we conclude the paper in Section 7.

## 2. OVERVIEW ON IMS SERVICE AUTHENTICATION

Third Generation Partnership Project (3GPP) has provided the bootstrapping of application security to authenticate the subscriber by defining a Generic Bootstrapping Architecture (GBA) [9] based on Authentication and Key Agreement (AKA) protocol. The GBA model can be utilized to authenticate subscriber before accessing multimedia services and applications over HTTP. The GBA consists of five entities [10]:

- The UE (User Equipment): is a UICC (Universal Integrated Circuit Card) containing USIM or ISIM related information that supports HTTP Digest AKA (Authentication & Key Agreement)
- The BSF (Bootstrapping Server Function): participates in the GBA through mutually authenticating with the UE using the AKA protocol, and agreeing on session keys that are afterwards applied between UE and the NAF.
- The NAF (Network Authentication Function): has the functionality of locating and communicating securely with subscriber's BSF
  - The HSS (Home Subscriber Server): is the master database of IMS that stores IMS user profiles.

### 2.1. Subscribers Identification in IMS

There are two types of identities that are associated with an IMS user.

1. A private identity (IMPI), where every IMS user shall have one or more Private User Identities, but there is only one private user identity stored in each ISIM card. The private identity is assigned by the home network operator, and is used for registration, authentication, authorization, administration, and accounting purposes. The private user identity takes the form of a Network Access Identifier (NAI).
   2. A public identity (IMPU) where every IMS user shall have one or more Public User Identities and each ISIM card stores at least one public user identity. The Public User Identities are used by any user for requesting communications to other users and to services access. The Public User Identity takes the form of a SIP URI or the "tel:"-URI format.

### 2.2. Bootstrapping Authentication Procedure

In order to allow the access to services over HTTP in a secure manner, IMS uses the Generic Bootstrapping Architecture (GBA) [9]. The GBA performs authentication between the Bootstrapping Server Function (BSF) and the UE, which is also based on AKA. Figure 1 shows the GBA authentication for services access in IMS.

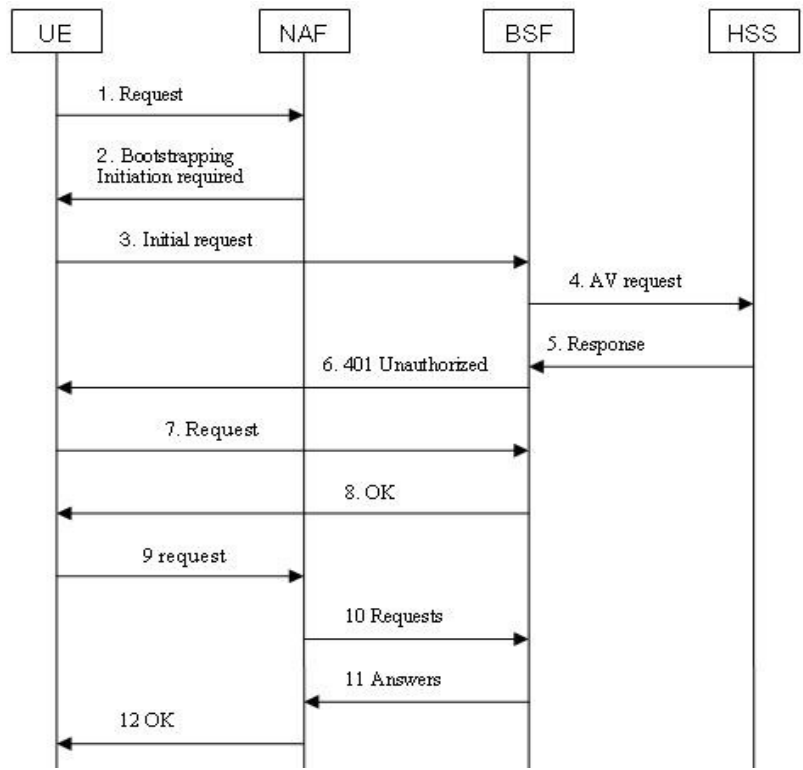

Figure 1:  GBA Authentication

Before the service access, the UE communicates with the NAF to verify if the bootstrapping procedure is needed (message 1). If it is the case and if there are no available bootstrapping parameters in the UE (message 2), the UE will contact the BSF by sending an HTTP request including the private user identity (message 3). Then, the BSF contacts the HSS to get the GBA User Security Setting (GUSS) and the Authentication Vector (AV) which includes RAND, AUTN, CK, IK and XRES (messages 4-5).

In order to demand the UE to authenticate itself, the BSF then forwards the RAND and AUTN to the UE in a "401 (Unauthorized) message" without including the CK, IK and XRES (message 6). The UE then checks AUTN (following the same procedure described in IMS authentication) to verify that the challenge is from an authorized network; and calculates CK, IK and RES. This will result in session keys IK and CK in both BSF and UE.

The UE sends another HTTP request to the BSF, containing the Digest AKA response which is calculated using RES (message 7).

The BSF authenticates the UE by verifying the Digest AKA response and generates key material Ks by concatenating CK and IK. The BSF also generates B-TID (Bootstrapping Transaction Identifier) and sends a "200 OK message", including the B-TID to the UE to indicate the success of the authentication (message 8).

The UE uses CK and IK to calculate the key material Ks and both the UE and BSF use the Ks to derive the key material Ks-NAF. The UE contacts the NAF and provides the B-TID and a digest calculated using Ks_NAF (message 9). The NAF then requests the corresponding Ks_NAF and GUSS from the BSF by sending B-TID (message 10).

After receiving the BSF response (message 11), the NAF calculates the digest values using Ks_NAF and compares the calculated values with the received ones from the UE to be able to authenticate the UE (message 12).

We notice that the BSF is the entity which communicates with all the other entities and has the important role to verify the UE's signatures. In real cases, there will be simultaneous UEs requesting for authentication and this will lead to a potential bottleneck of signature verification at the BSF.

## 3. Overview On Identity Based Cryptography IBC

The Identity Based cryptography (IBC) has emerged as a long-term evolution or substitution to Public Key Infrastructure PKI. It is a cryptosystem in which the public key is retrieved from an identity of the entity (user) and the private key is more precisely the public key multiplied by the secret key of the server. The latter is responsible of the private key distribution and is called the Private Key Generator (PKG).

The IBC concept is old and was first proposed by Shamir in 1984 [11], then the first fully practical and secure identity-based public key encryption scheme was presented by D. Boneh and M. Franklin in [8], using the fundamental operations of Elliptic Curve Cryptography (ECC). Since then, a rapid development of Identity based cryptosystems has taken place.

The IBC is based on the bilinear pairing which is presented later.

Shamir's original motivation for identity-based encryption was to simplify the certificate management in e-mail systems. This solution is illustrated in Figure 2.

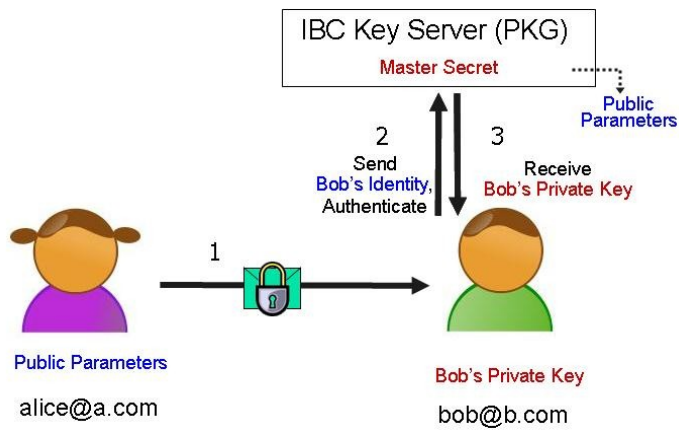

Figure 2: IBC in mail scenario

### 3.1. Bilinear Pairings

Let $(G_1; +)$ and $(G_2; .)$ be two cyclic groups of prime order q. $G_1$ is an additive group and $G_2$ as a multiplicative group. The bilinear pairing is given as e: $G_1 * G_1 \rightarrow G_2$, which satisfies the following properties:

1. Bilinearity:
   For all P; Q, R $\in$ $G_1$; e(P + Q, R) = e(P, R). e(Q, R) and e(P, Q + R) = e(P, Q). e(P, R);

2.  Non-degeneracy: There exists P; Q $\in$ G$_1$ such that e(P, Q) $\neq$ 1;

3.  Computability: It is efficient to compute e(P, Q) $\triangledown$P; Q $\in$ G$_1$.

A bilinear map satisfying the three properties above can be considered as an admissible bilinear map [8].

## 3.2. IBC basic Functions

In IBC, the public key is usually related to the public identity of the user (for example, email addressee). The following functional aspects are always present in any IBC cryptosystem.

1.  *System Setup*: IBC systems rely on a trusted central authority that manages the parameters with which keys are created. This authority is called the Private Key Generator (PKG). The PKG creates its parameters, including a master secret S used to generate the private keys for users. The system parameters are: the order q, the prime number p, the generator point P, PKG public point Ppub= S.P and the hash functions.

2.  *Encryption*: When a user (Alice) wishes to send an encrypted message to another user (Bob), she encrypts the message to him by computing or obtaining the public key, Kpub$_{Bob}$, and then encrypting a plaintext message M with Kpub$_{Bob}$ to obtain cipher message C.

3.  *Key Extraction*: When Bob receives the message, he wants to decrypt it. He authenticates himself to the PKG and obtains the secret key Kpriv$_{Bob}$ that he uses to decrypt the cipher message C.

4.  *Decryption*: When Bob receives Kpriv$_{Bob}$, he decrypts the cipher message C to obtain the plaintext message M.

## 3.3. Batch Verification Scheme

In some architectures we need to verify a signature to authenticate the users, where the party concerned by the verification presents a bottleneck in the system if the load is high (case of many users asking for authentication). One solution is the Batch verification which consists on the verification of all the signatures received in a time window with rather short time compared to verifying each signature one after the other. The batch cryptography based on RSA was introduced by Fiat [12] in 1989. Some other batch signature schemes were proposed later like [13].

Zhang et al [14] used 3 pairing operations to verify a single signature. To verify n signatures, they needed 3 pairing operations instead of 3n pairing operations. In other words, the verification time of the dominant operation (i.e., pairing) is independent of the number of signatures to verify [14]. As a result, the time spent on verifying a large number of signatures will be decreased.

## 4. PROPOSED IMS SERVICE AUTHENTICATION BASED ON IBC

In this section, we describe our novel solution which uses IBC in the IMS GBA authentication to generate the private keys of the UE instead of using AKA mechanisms. Our objective is to personalize the IMS service authentication process through using IMPU. This new scheme will lead to better performances when there are simultaneous authentication requests thanks to using Identity-based Batch Verification.

The HSS has a PKG server which has the role to generate the private keys for the UE. We use bilinear map e: e: G1 * G1 $\rightarrow$ G2. The PKG randomly chooses s1, s2 $\in$ $Z^*_q$ as its two master keys, and computes Ppub$_1$ = s$_1$.P, Ppub$_2$ = s$_2$.P as its public keys.

In our work, we assume that the UE has the shared key sk with HSS and the PKG parameters (order q, prime number p, P, $Ppub_1$, $Ppub_2$ and MapToPoint function) stored in the ISIM card.

We present in Table 1, all the notation used in the solution.

Table 1. Notations

| Notation | Descriptions |
| --- | --- |
| $UE^i$ | The ith UE: User Equipment |
| $G_1$ | A cyclic additive group |
| $G_2$ | A cyclic multiplicative group |
| P | The generator of the cyclic additive group $G_1$ |
| e | A bilinear map: $G_1 \times G_1 \rightarrow G_2$ |
| q | The order of the group $G_1$ |
| r | A random nonce |
| $s_1, s_2$ | The private master keys of the PKG |
| $Ppub_1, Ppub_2$ | The public keys of the PKG, $Ppub_1 = s_1.P$ and $Pub_2 = s_2.P$ |
| H(.) | A MapToPoint hash [8] function such as H : $\{0, 1\}^* \rightarrow G$ |
| UEID | UEID = H(IMPU) |
| $Kpub^i_1, Kpub^i_2$ | The public keys of the $UE^i$ |
| $SK^i_1, SK^i_2$ | The private keys of $UE^i$, $SK^i_1 = s_1.Ppub_1$ and $SK^i_2 = s_2.Ppub_2$ |
| $RAND^i$ | Random value to authenticate $UE^i$ |
| h(.) | A one-way hash function such that SHA-1 |
| ‖ | Message concatenation operation, which appends several messages together in a special format |

## 4.1. Description of the solution

The proposed solution is explained in the following steps. In Figure 3, we illustrate all the messages exchanged to authenticate the ith UE ($UE^i$).

**Step 1.** (messages 1 and 2) $UE^i$ starts communication with the NAF without GBA parameters. If the NAF requires the use of shared keys obtained by means of the GBA, but the request from the UE does not include GBA-related parameters, the NAF replies with a bootstrapping initiation message.

**Step 2.** (messages 3, 4 and 5) $UE^i$ sends a HTTP request to the BSF (Bootstrapping Server Function) including its IMS private user identity ($IMPI^i$) and public user identity ($IMPU^i$). The BSF then retrieves from the HSS:

1. the public keys $Kpub^i_1$ and $Kpub^i_2$ (generated using $IMPU^i$) from the PKG.

$$Kpub^i_1 = r \cdot P \quad \text{and} \quad Kpub^i_2 = UEID \oplus H(r.Ppub_1)$$

where r is a random number, $\oplus$ XOR operation and UEID = H(IMPU$^i$)

2. the complete set of GBA user security settings (GUSS$^i$),
3. an Authentication Vector (AV$^i$) containing the RAND$^i$ and PKG parameters,
4. the private keys Kpriv$^i_1$ and Kpriv$^i_2$ encrypted with shared key sk where

$$Kpriv^i_1 = s_1 \cdot Kpub^i_1 \quad \text{and} \quad Kpriv^i_2 = s_2 \cdot H(Kpub^i_1 \parallel Kpub^i_2)$$

**Step 3.** (message 6) In order to demand the UE$^i$ to authenticate itself, the BSF forwards Kpub$^i_1$, Kpub$^i_2$, [Kpriv$^i_1$]sk and [Kpriv$^i_2$]sk and RAND$^i$ to the UE in the "401 (Unauthorized) message".

**Step 4.** (message 7) The UE$^i$ extracts its private keys Kpriv$^i_1$ and Kpriv$^i_2$ using the shared key sk which is stored in the ISIM card. Then, the UE$^i$ hashes the RAND$^i$ and computes the signature Sig$^i_1$ where: $\qquad Sig^i_1 = Kpriv^i_1 + h(RAND^i) \cdot Kpriv^i_2$

The UE$^i$ sends IMPU$^i$, RAND$^i$, Sig$^i_1$ to the BSF in an HTTP request. To verify the UE"s signature, the BSF has already the PKG parameters and Kpub$^i_1$ and Kpub$^i_2$ corresponding to IMPU$^i$. Sig$^i_1$ is valid if

$$e(Sig^i_1, P) = e(Kpub^i_1, Ppub_1) \cdot e(h(RAND^i) \cdot H(Kpub^i_1 \parallel Kpub^i_2), Ppub_2) \qquad (1)$$

If the verification phase is successful, then, the user is authenticated. This is how equation 1 is computed:

$$e(Sig^i_1, P) = e(Kpriv^i_1 + h(RAND^i) \cdot Kpriv^i_2, P)$$

$$= e(Kpriv^i_1, P) \cdot e(h(RAND^i).Kpriv^i_2, P)$$

$$= e(s_1.Kpub^i_1, P) \cdot e(h(RAND^i).s_2.H(Kpub^i_1 \parallel Kpub^i_2), P)$$

$$= e(Kpub^i_1, s_1.P) \cdot e(h(RAND^i).H(Kpub^i_1 \parallel Kpub^i_2), s_2.P)$$

$$= e(Kpub^i_1, Ppub_1) \cdot e(h(RAND^i).H(Kpub^i_1 \parallel Kpub^i_2), Ppub_2)$$

**Step 5.** (message 8) After the successful verification, the BSF generates B-TID$^i$ (Bootstrapping ID) and stores it with the IMPU$^i$ and GUSS$^i$. Then, the BSF sends to the UE a "200 OK message" including the B-TID$^i$ encrypted with UE$^i$'s public key kpub$^i_1$ (BSF can use any asymmetric elliptic curve algorithm). After receiving the message, the UE retrieves the B-TID using Kpriv$^i_1$.

In our solution, there is no key material Ks stored in the UE and the BSF. Our system is based on asymmetric cryptography. The shared key sk between the UE$^i$ and the HSS is used to encrypt the UE$^i$'s Kpriv$^i_1$ and Kpriv$^i_2$. The BSF cannot retrieve these keys and has to encrypt B-TID$^i$ using UE$^i$'s kpub$^i_1$.

**Step 6.** (message 9) The following steps apply Elliptic Curve Diffie-Hellman (ECDH) Protocol. This key agreement protocol will be used to generate the Ks-NAF key. The UE$^i$ and the NAF first have to agree whether to use the shared keys obtained by means of the GBA. The UE$^i$ chooses a random value 'a' to generate 'a. Kpub$^i_1$' and provide the IMPU$^i$, B-TID$^i$, 'a. Kpub$^i_1$', and a signature of B-TID$^i$ to allow it to retrieve the corresponding keys from the BSF. The Signature of BTID$^i$ is: $\qquad Sig^i_2 = Kpriv^i_1 + h(B\text{-}TID^i) \cdot Kpriv^i_2$

**Step 7.** (message 10) The NAF sends to the BSF the NAF-ID, the IMPU$^i$, B-TID$^i$ and Sig$^i_2$ to request for GUSS$^i$ and PKG parameters. NAF-ID is used by the BSF to verify that the NAF is authorized to use that hostname.

**Step 8.** (message 11) First of all, the BSF verifies the signature using Kpub$^i_1$ and Kpub$^i_2$ (same verification as in step 4, equation (1)). Then, it retrieves the GUSS$^i$ and PKG parameters using

B-TID$^i$ and IMPU$^i$. Finally, it supplies to the NAF the IMPU$^i$, Kpub$^i_1$ and Kpub$^i_2$, GUSS$^i$, and the PKG parameters.

**Step 9.** (message 12) The NAF checks the authentication and the authorization of the UE$^i$ to the services according to the received GUSS$^i$ and then generates a random value 'b' and send to the UE$^i$ 'b. Kpub$^i_1$'. After receiving the message, the UE$^i$ and the NAF will have the same Ks-**NAF = a.b.Kpub$^i_1$**. Once the execution of the protocol is completed, the UE$^i$ and the NAF will communicate in a secure way and UE$^i$ will be granted the services.

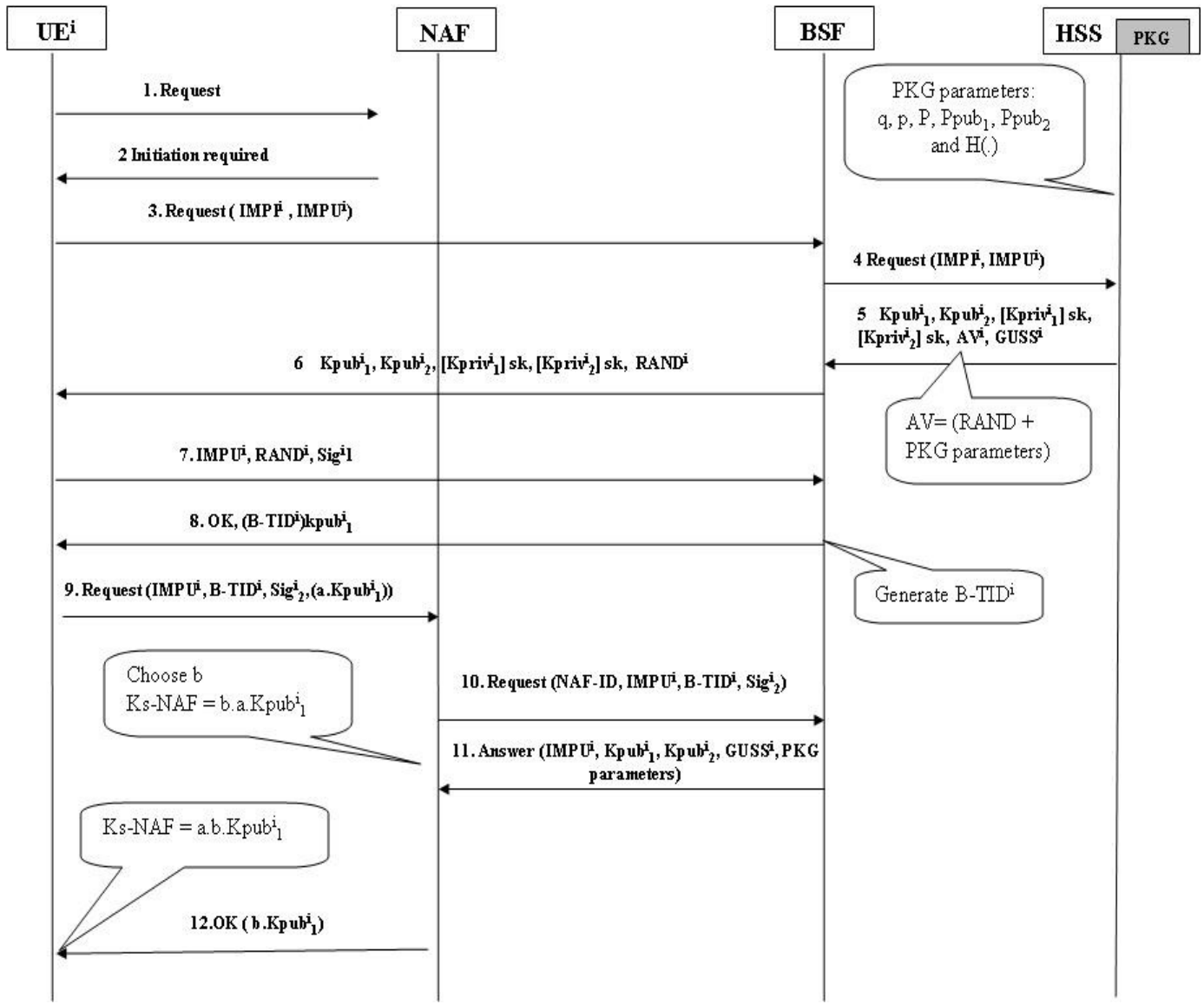

Figure 3. IMS Service Authentication for the ith UE

## 4.4. Batch Verification

To verify Sig$^i_1$ and Sig$^i_2$, the BSF needs one MapToPoint hash (H), one multiplication, and three pairing operations. We estimate that the computation cost of a pairing operation is much higher than the cost of a MapToPoint hash and a multiplication operation.

We suppose that we have n UEs which belong to the same HSS and communicate through the same BSF. The latter will receive $(IMPU^1, RAND^1, Sig^1_j)$, $(IMPU^2, RAND^2, Sig^2_j)$, ..., $(IMPU^n, RAND^n, Sig^n_j)$, respectively, which are sent by n distinct UEs $UE^1$, $UE^2$, ... $UE^n$ and j= 1 or 2.

We just focus in this paragraph on j = 1 because it is respectively the same for j = 2. All the signatures, denoted $Sig^1_1$, $Sig^2_1$, ..., $Sig^n_1$, are valid if

$$e(\sum_{i=1}^{n} Sig^i_1, P) = e(\sum_{i=1}^{n} Kpub^i_1, Ppub_1) . e(\sum_{i=1}^{n} h(RAND^i). H(Kpub^i_1 \| Kpub^i_2), Ppub_2)$$

$$(2)$$

We proceed like this to give the details of equation 2:

$$e(\sum_{i=1}^{n} Sig^i_1, P) = e(\sum_{i=1}^{n} (Kpriv^i_1 + h(RAND^i) . Kpriv^i_2), P)$$

$$= e(\sum_{i=1}^{n} Kpriv^i_1, P) . e(\sum_{i=1}^{n} h(RAND^i).Kpriv^i_2, P)$$

$$= e(\sum_{i=1}^{n} s_1.Kpub^i_1, P) . e(\sum_{i=1}^{n} h(RAND^i).s_2.H(Kpub^i_1 \| Kpub^i_2), P)$$

$$= e(\sum_{i=1}^{n} Kpub^i_1, s_1.P) . e(\sum_{i=1}^{n} h(RAND^i).H(Kpub^i_1 \| Kpub^i_2), s_2.P)$$

$$= e(\sum_{i=1}^{n} Kpub^i_1, Ppub_1) . e(\sum_{i=1}^{n} h(RAND^i).H(Kpub^i_1 \| Kpub^i_2), Ppub_2)$$

## 4.2. Advantages of the proposed solution

Our proposed solution has a number of advantages as follows:

1. The applied batch verification can dramatically reduce the verification delay, particularly when verifying a large number of signatures. From the batch verification equation (2), the computation cost that the BSF spends on verifying n signatures is dominantly comprised of n MapToPoint hash, n multiplication, 3n addition, n one-way hash, and 3 pairing operations. Without the batch verification we will have 3n pairing operations to realize.

2. Our proposed solution doesn't use AKA-MD5 in the GBA and provides more simplicity since it is based on Elliptic Curve Cryptography (ECC). Therefore, it is more efficient and preferable in the applications that require low memory and rapid transaction. Furthermore, for elliptic-curve-based protocols, it is assumed that finding the discrete logarithm of an elliptic curve element is infeasible.

3. Our system is based on asymmetric cryptography, where the shared key ks between the UE and the HSS is used to encrypt the UE's private keys. We don't need to have mutual authentication with the BSF. The latter cannot retrieve these keys and has to encrypt B-TID using UE's $Kpub_1$. Then, even if there is no mutual authentication between the UE and the BSF, security will be always guaranteed.

## 5. SECURITY ANALYSIS

The UE based frauds lead to illegitimate use of IMS services with stolen credentials. In the following, we analyze the Eavesdropping attack and we show the robustness of our proposed solution against this attack. In our point of view, it is the most common attack on IMS. We also explain how the message authentication is secure. All the messages used in this section are the ones illustrated in Figure 3.

## 5.1. Eavesdropping Attack

During the authentication phase (messages 3 to 8), a malicious user can play the role of a Man in the Middle (MITM), listens to the communication and retrieves the GUSS related to the IMPI and IMPU of the legitimate user. The malicious user then tries to connect to the system using his ISIM card (containing his private identity IMPI' and the legitimate's IMPU). The HSS will reject the request because it doesn't have the couple (IMPI', IMPU) in its database.

## 5.2. Message Authentication.

In the proposed scheme, the signature $\mathbf{Sig_1 = Kpriv_1 + h(RAND) \cdot Kpriv_2}$ is actually a one-time identity based signature (for each authentication, the HSS chooses a different random number r). Without knowing the private key $Kpriv_1$ and $Kpriv_2$, it is infeasible to forge a valid signature. Because of the NP-hard computation complexity of Diffie-Hellman problem in G, it is difficult to derive the private keys $Kpriv_1$ and $Kpriv_2$ by using $Kpub_1$, $Ppub_1$, P, and $H(Kpub^i_1 \parallel Kpub^i_2)$. At the same time, because $Sig_1$ or $Sig_2$ is a Diophantine equation [14], by only knowing $Sig_1$ and $h(RAND)$ (or $Sig_2$ and $h(B-TID)$), it is still difficult to get the private keys. Therefore, the one-time identity-based signature is unforgeable, and the property of message authentication is fully satisfied.

## 6. Performance Evaluation

In this section, we evaluate the performance of the novel proposed solution especially for verifying the delay caused by the BSF. We believe that IBC performance is the most critical point that can judge the feasibility of deploying our solution, and the measures that we got are not discouraging especially with the use of Batch Verification.

For the whole scenario, the most important parameter influencing the performance will be the speed of the cryptographic operations (such as private/public key pair generation, encryption/decryption, and signature/verification time). In this performance analysis, we mainly measure the speed of the cryptographic operations, without considering the underlying network architecture at this phase. We estimate that the BSF will be the bottleneck of the service authentication process since it processes many HTTP messages and has the important role to verify the UE's signatures.

We notice the existence of IMS platforms [15] however without any implementation of the GBA Authentication for services access that why we need to implement our own testbed. We implemented the PKG, which is found within the HSS, through employing the IBE demo provided within Miracl library [16]. We use the elliptic curve $y^2 = x^3 + 1 \bmod p$ where p is 256 bits prime number.

In the following, all the measures are real measures from the implementation realized using an Intel® Core ™2 CPU T5470 @ 1,60GHZ. We observed that the time needed to generate PKG parameters (160 bits q, 256 bits q, 512 bits point P, 512 Point Ppub, 160 bits secret S and 512 bits cube root of unity in Fp2) is around 14 ms. To generate $Kpub_1$ and $Kpub_2$, we use a MapToPoint function, which has the role of finding a point in the curve corresponding to the Hash of the IMPU. We found that the time needed for MapToPoint $T_{mtp}$ will be in the order of 4,4 ms. To generate $Kpriv_1$ and $Kpriv_2$, the PKG needs almost 7,5ms. The time needed for bilinear pairing $T_{bp}$ is about 9,3 ms and the time for multiplication $T_{mul}$ is about 1,5 ms.

For the verification of $Sig_1$ or $Sig_2$, we need 3 bilinear pairings, 1 MapToPoint and 1 multiplication, so for 1 person $T_v = 3\ T_{bp} + T_{mtp} + T_{mul} \approx 33,8$ ms. The BSF has a maximum capacity, we note as N. If we have more than N UEs simultaneously requesting authentication, the system will reject or delay the answer. With the Batch Verification, this can be avoided since the verification for n signature will cost 3 bilinear pairings, n MapToPoint and n multiplication. Table 2 shows the time needed that we deduced for different number of UEs (we choose N >=1000 UEs).

From the performance point of view, the asymmetric system seems to need more time to finish all operations than symmetric one but this is not harmful to our work. Furthermore, from the security point of view, the identity Based Cryptography IBC is more secure (using 160 key instead of 128 key for AKA) and we calculated that we need about 4 min to authenticate 50000 UEs, and it seams to be an encouraging result.

Table.2  Time needed to authenticate all the UEs

| Verification method Number of UEs | Without Batch Verification (s) | With Batch Verification |
|:---:|:---:|:---:|
| 1000 | 33,8s | ~ 5,9 s |
| 5000 | 169 s | ~ 29,5 |
| 10000 | 338 s | ~ 59 s |
| 50000 | 1690 s | ~ 295 s |

# 7. CONCLUSION

IP Multimedia Subsystem (IMS) merges the internet with the cellular world to provide ubiquitous access to internet technologies and to provide consumers with appealing services. IMS authentication falls short in one hand to be realized in a personalized manner, which is an important prerequisite in new services such as social internet ones. On the other hand, using AKA in IMS proved to have some weakness, like short key for cryptographic purposes. In this paper, we integrate the Identity Based Cryptography (IBC) in the IMS Service Authentication scheme. Since IBC is based on Elliptic Curve Cryptography (ECC), it is more efficient and preferable in the applications that require low memory and rapid transaction. Security is assured thanks to using of symmetric protocol with a shared key (ks) between the UE and the HSS, an asymmetric protocol for signature, and Diffie-Hellman for key agreement. We define a Batch Verification on the Bootstrapping Server Function BSF to decrease verification delay and the authentication response time. We analyzed some security aspects of the solutions and we showed that our proposed solution can prevent against the attacks. To validate the performance of our proposed solution, we implemented the cryptographic operation in our proposed solution including the IBC procedures. We observe that the use of asymmetric cryptographic procedures will lead to longer running time than symmetric procedures. However, the Batch Verification helps the BSF to verify the UEs signature in a reasonable time. As a future work, we need to improve our testbed by integrating it to an IMS core for a complete authentication session and to perform tests in realistic network conditions.

**Authors**

**Mohamed Abid** is currently a Phd Student in third year (Start in 2006) in TELECOM & Management Sudparis (France). He holds a Master Diploma in Computer science in 2005 from ENSI Ecole Nationale des Sciences de l'Informatique (Tunisia). His research interests include control access in networks (network security), Biometric and cryptography, Future Internet, and Handover in wireless networks.

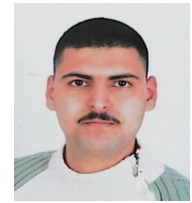

**Songbo Song** is currently a Phd Student in first year (Start in 2008) in TELECOM & Management Sudparis and Telecom R&D (Orange Labs), Issy Les Moulineaux (France). He holds an Engineer Diploma in Telecommnuication in 2008 from TELECOM & Management Sudparis. His research interests include Personnalized access in IMS architecture and context awerness.

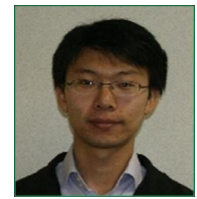

**Hassnaa Moustafa** is a Senior Research Engineer at France Telecom R&D (Orange Labs), Issy Les Moulineaux (France) since January 2005. She obtained her PhD in Computer and Networks from Telecom ParisTech in December 2004 and her Master degree in distributed systems in September 2001. Her research interests include mobile networks basically ad hoc networks and vehicular networks, mainly routing, security, authentication and access control are part of her research interests in these types of networks. Moreover, she is interested in NGN, IPTV, services' convergence and personalization. She is a regular member in the IETF and a member of the IEEE.

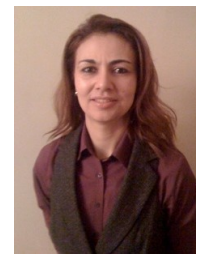

**Hossam Afifi** is a Professor in computer science and communications, of Institut TELECOM & Management Sudparis, France. (Security and Mobility Group) from September 2002. He earned his Tenure in Computer Science (HDR) in September 1999 from University of Rennes I, (France).on Protocol Design and Optimization. And earned his doctorale in computer science in June 1993 from INRIA, Sophia-Antipolis, (France) on Dynamic Message Routing using Directory Service based procedures. From 2001, he is a member of the Security and Mobility Group in Institut TELECOM & Management Sudparis, France. His current research interests include security evaluation, Multimedia Services, Mobility in networks and Next Generation Network NGN

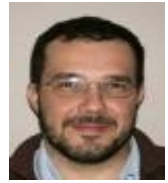